\pdfoutput=1

\documentclass[11pt]{article}

\usepackage[]{acl}

\usepackage{times}
\usepackage{latexsym}

\usepackage[T1]{fontenc}

\usepackage[utf8]{inputenc}

\usepackage{microtype}

\usepackage{inconsolata}

\usepackage{graphicx}
\usepackage{booktabs, multirow} 
\usepackage{soul}
\usepackage{xcolor,colortbl} 
\usepackage{changepage,threeparttable} 
\usepackage{graphicx}
\usepackage{amsmath}
\usepackage[capitalise,nameinlink]{cleveref}
\usepackage{tcolorbox}
\usepackage{wrapfig}

\newcommand{\rowhighlight}{\cellcolor[gray]{0.9}}

\title{Document Attribution:\\Examining Citation Relationships using Large Language Models}

\author{ 
 \textbf{Vipula Rawte\textsuperscript{1}\thanks{Work done while the first author was an intern at Adobe Research}},
 \textbf{Ryan Rossi\textsuperscript{2}},
 \textbf{Franck Dernoncourt\textsuperscript{2}},
 \textbf{Nedim Lipka\textsuperscript{2}}
\\
\\
 \textsuperscript{1}Adobe Inc.
 \textsuperscript{2}Adobe Research
\\
 \small{
    \href{mailto:vrawte@adobe.com}{vrawte@adobe.com}
 }
}

\begin{document}
\maketitle
\begin{abstract}

As Large Language Models (LLMs) are increasingly applied to \textbf{document-based tasks} - such as document summarization, question answering, and information extraction - where user requirements focus on retrieving information from provided documents rather than relying on the model's parametric knowledge, ensuring the trustworthiness and interpretability of these systems has become a critical concern. A central approach to addressing this challenge is \emph{attribution}, which involves tracing the generated outputs back to their source documents. However, since LLMs can produce inaccurate or imprecise responses, it is crucial to assess the reliability of these citations. 

To tackle this, our work proposes two techniques. (1) A \textbf{zero-shot} approach that frames attribution as a straightforward textual entailment task. Our method using \texttt{flan-ul2} demonstrates an improvement of 0.27\% and 2.4\% over the best baseline of ID and OOD sets of AttributionBench \citet{li-etal-2024-attributionbench}, respectively. (2) We also explore the role of the \textbf{attention mechanism} in enhancing the attribution process. Using a smaller LLM, \texttt{flan-t5-small}, the F1 scores outperform the baseline across almost all layers except layer 4 and layers 8 through 11.
\end{abstract}

\section{Introduction}

Attribution in Large Language Models refers to tracing the origins of information embedded in the model's outputs. This involves identifying the specific datasets, documents, or text segments contributing to the generated response. Attribution is essential for verifying the information's provenance, ensuring the generated content's accuracy and reliability, and addressing concerns regarding plagiarism, accountability, and transparency in AI systems. Attribution methods typically involve mapping responses to the relevant data sources that influenced the model’s generation.

In LLMs, attribution systematically links the model’s outputs to their source materials, facilitating the identification of the exact documents, datasets, or references that informed the generated response. The primary goal is to uphold transparency, validate factual correctness, and give proper credit to sources. This process is critical for maintaining the credibility and accountability of generative AI systems.

Attribution methods are fundamental for enhancing the interpretability and dependability of LLMs. They support the model’s output by providing citations or references, improving accuracy, and reducing the risk of misinformation. This ensures that each response is substantiated by relevant evidence, forming a basis for assessing the sufficiency and relevance of the underlying data.

Research on LLM attribution methodologies encompasses citation generation, claim verification, and hallucination detection techniques. These strategies are aimed at improving the quality and reliability of LLM-generated content. However, challenges remain in implementing adequate attribution, including the need for robust validation mechanisms, managing cases where sources influence the model’s reasoning indirectly, handling structured data or non-textual sources (e.g., tables, figures, or images), and addressing the complexities of multi-lingual or cross-lingual data. Overcoming these challenges is essential for successfully integrating attribution methods within LLMs.

As AI and machine learning systems become increasingly prevalent, the demand for accountability, transparency, and reliability intensifies. Attribution techniques are pivotal in achieving these objectives, positioning them as a key area of research and development to advance AI technologies and ensure their responsible deployment. 

The main \textbf{contributions} of this work are:

\begin{itemize}
    \item A simple zero-shot prompting technique following the idea of textual entailment.

    \item An attention-based binary classification technique exploring whether attention could help achieve the attribution better.
    
\end{itemize}

\section{Related Work}

Attribution in LLMs has become a vital research area focused on tracing content origins and ensuring accuracy and accountability. Key studies have introduced various techniques and addressed challenges in this field.

\citet{pasunuru2023eliciting} propose a minimal-supervision method for eliciting attributions, improving scalability, and reducing the need for extensive human input. An interactive visual tool for attribution is introduced \citet{lee2024llm}, aiming to enhance transparency by making attributions more accessible to non-technical users. \citet{zhou-etal-2024-explaining} explore attribution in low-resource settings, emphasizing its potential to explain model behavior when data and resources are limited.

The Captum interpretability library is used in \citet{miglani-etal-2023-using} for generative LLMs, offering insights into the factors influencing model predictions. \citet{khalifa2024sourceaware} argue that source-aware training enhances attribution by linking knowledge to specific sources, improving content reliability. The issue of false attribution, stressing the need for more accurate methodologies, is highlighted in \citet{adewumi2024limitations}.

\citet{bohnet2023attributed} focus on attribution in question-answering systems, proposing methods for evaluating and modeling attributions in QA contexts.  A survey of LLM attribution research, summarizing key techniques, challenges, and developments, is provided in \citet{li2023survey}. Lastly, \citet{yue-etal-2023-automatic} explores the automated evaluation of attribution, aiming to streamline validation processes in practical applications.


\begin{table*}[!]
\centering
\scriptsize
\resizebox{\textwidth}{!}{%
\begin{tabular}{ll|ccc|ccc|ccc|ccc|c}
\toprule
\textbf{Setting } &\textbf{Model (Size)} &\multicolumn{3}{c}{\textbf{ExpertQA (\#=612)}} &\multicolumn{3}{c}{\textbf{Stanford-GenSearch (\#=600)}} &\multicolumn{3}{c}{\textbf{AttributedQA (\#=230)}} &\multicolumn{3}{c}{\textbf{LFQA (\#=168)}} &\textbf{ID-Avg.} \\
\cmidrule(lr){3-5} \cmidrule(lr){6-8} \cmidrule(lr){9-11} \cmidrule(lr){12-14}
& &\textbf{F1 ↑} &\textbf{FP ↓ } &\textbf{FN ↓ } &\textbf{F1 ↑} &\textbf{FP ↓ } &\textbf{FN ↓ } &\textbf{F1 ↑} &\textbf{FP ↓ } &\textbf{FN ↓ } &\textbf{F1 ↑} &\textbf{FP ↓ } &\textbf{FN ↓ } &\textbf{F1 ↑} \\
\toprule

\multirow{11}{*}{\textbf{Zero-shot \citet{li-etal-2024-attributionbench}}} 
& \textbf{FLAN-T5 (770M)} & \rowhighlight 38.2 & \rowhighlight 1.3 & \rowhighlight 47.4 & \rowhighlight 73.5 & \rowhighlight 15 & \rowhighlight 11.5 & \rowhighlight 80.4 & \rowhighlight 12.2 & \rowhighlight 7.4 & \rowhighlight 37.2 & \rowhighlight 0 & \rowhighlight 48.2 & \rowhighlight 57.3 \\
& \textbf{FLAN-T5 (3B)} & 55.6 & 15.8 & 27.9 & 74 & 17.2 & 8.7 & 79.8 & 15.2 & 4.8 & 75.3 & 6.5 & 17.9 & 71.2 \\
& \textbf{AttrScore-FLAN-T5 (3B)} & \rowhighlight 55.7 & \rowhighlight 32.4 & \rowhighlight 9.6 & \rowhighlight 64.6 & \rowhighlight 27.3 & \rowhighlight 6.5 & \rowhighlight 80.5 & \rowhighlight 16.5 & \rowhighlight 2.6 & \rowhighlight 71.4 & \rowhighlight 21.4 & \rowhighlight 6.5 & \rowhighlight 68.1 \\
& \textbf{FLAN-T5 (11B)} & 52 & 36.4 & 7.5 & 59.2 & 32.7 & 5 & 78.6 & 18.3 & 2.6 & 79.8 & 10.1 & 10.1 & 67.4 \\
& \textbf{T5-XXL-TRUE (11B)} & \rowhighlight 54.5 & \rowhighlight 17.8 & \rowhighlight 27.3 & \rowhighlight 68.5 & \rowhighlight 16.2 & \rowhighlight 15.3 & \rowhighlight \textbf{85.2} & \rowhighlight 7.8 & \rowhighlight 7 & \rowhighlight 80.4 & \rowhighlight 1.2 & \rowhighlight 17.9 & \rowhighlight 72.2 \\
& \textbf{FLAN-UL2 (20B)} & 59.4 & 22.5 & 18 & 72.5 & 19.2 & 8 & 82.5 & 13 & 4.3 & 80.1 & 4.2 & 15.5 & 73.6 \\
& \textbf{AttrScore-Alpaca (7B)} & \rowhighlight 47.4 & \rowhighlight 11.1 & \rowhighlight 37.7 & \rowhighlight 68.6 & \rowhighlight 21.2 & \rowhighlight 9.8 & \rowhighlight 79 & \rowhighlight 14.8 & \rowhighlight 6.1 & \rowhighlight 68.7 & \rowhighlight 10.1 & \rowhighlight 20.8 & \rowhighlight 65.9 \\
& \textbf{GPT-3.5 (w/o CoT)} & 55.3 & 30.4 & 12.1 & 62 & 30.5 & 3.8 & 74.7 & 20.9 & 3.5 & 72.6 & 22 & 4.2 & 66.2 \\
& \textbf{GPT-3.5 (w/ CoT)} & \rowhighlight \textbf{60.4} & \rowhighlight 23 & \rowhighlight 16.2 & \rowhighlight 66.1 & \rowhighlight 25.5 & \rowhighlight 7.2 & \rowhighlight 78.9 & \rowhighlight 14.3 & \rowhighlight 6.5 & \rowhighlight 73.4 & \rowhighlight 19.6 & \rowhighlight 6.5 & \rowhighlight 69.7 \\
& \textbf{GPT-4 (w/o CoT)} & 56.5 & 32.8 & 8 & 59.8 & 33.2 & 3.5 & 81 & 15.7 & 3 & 71.6 & 23.2 & 4.2 & 67.2 \\
& \textbf{GPT-4 (w/ CoT)} & \rowhighlight 59.2 & \rowhighlight 26.3 & \rowhighlight 13.9 & \rowhighlight 71.7 & \rowhighlight 19.5 & \rowhighlight 8.5 & \rowhighlight 82.2 & \rowhighlight 10 & \rowhighlight 7.8 & \rowhighlight 80.2 & \rowhighlight 14.9 & \rowhighlight 4.8 & \rowhighlight 73.3 \\
\midrule
\multirow{2}{*}{\textbf{Our Zero-shot}} 
& \textbf{gpt4-o (05-13-2024)} & 52 & 13.1 & 33 & 64.7 & \textbf{14} & 21.2 & 71.5 & 10 & 18.3 & 81.14 & 23.8 & 15.47 & 64.1 \\
& \textbf{flan-ul2 (20B)} & \rowhighlight 55 & \rowhighlight 32.7 & \rowhighlight 10 & \rowhighlight \textbf{75.2} & \rowhighlight 16 & \rowhighlight 8.7 & \rowhighlight 84.16 & \rowhighlight 20.86 & \rowhighlight 12.17 & \rowhighlight \textbf{85.38} & \rowhighlight 16.6 & \rowhighlight 13.09 & \rowhighlight \textbf{73.8} \\
\bottomrule
\end{tabular}%
}
\caption{We evaluate our zero-shot approach against the AttributionBench. Results highlighted in \textbf{bold} denote the highest performance. Our method performs better than existing approaches on the Stanford-GenSearch and LFQA sub-datasets. The average ID achieved using our method with \texttt{flan-ul2} is \textbf{73.8}, representing the highest value.}
\label{tab:results-id}
\end{table*}

\section{Method and Experimental Setup}

The attribution task defined in AttributionBench \citet{li-etal-2024-attributionbench} is framed as a binary classification problem, where the objective is to determine whether a given claim is attributable to its associated references. The work in AttributionBench explores this problem using both zero-shot inference and fine-tuning of LLMs. Similarly, our formulation adopts the same approach to the problem. However, we restrict our methodology to zero-shot experiments due to computational limitations. Additionally, we also investigate if attention layers could help improve the attribution.

\subsection{Zero-shot Textual Entailment}
We frame this attribution task as a textual entailment problem to ensure simplicity and efficiency.

\textbf{Textual entailment} refers to the relationship between two text fragments, typically a premise and a hypothesis, where the goal is to determine whether the premise entails the hypothesis. Formally, given two sentences \( S_1 \) (premise) and \( S_2 \) (hypothesis), textual entailment can be defined as a binary relation \( \text{Entail}(S_1, S_2) \), where:

\[
\text{Entail}(S_1, S_2) = 
\begin{cases} 
1, & \text{if } S_1 \text{ entails } S_2 \\
0, & \text{otherwise}
\end{cases}
\]

Here, \( S_1 \) entails \( S_2 \) if the meaning of \( S_1 \) logically supports or guarantees the truth of \( S_2 \). The task is to model this relation using techniques, such as deep learning models, to predict this entailment relationship based on large corpora of annotated text pairs.

\noindent \paragraph{Why zero-shot Textual Entailment?} The core challenge in zero-shot textual entailment is to build models that can generalize well to unseen tasks and relationships, relying purely on contextual understanding rather than task-specific fine-tuning. This is typically achieved through techniques like transfer learning, where models use their broad language understanding to handle specific inference tasks on the fly. For example, a model may be able to infer whether the statement ``It is raining outside'' entails ``The ground is wet'' without having been specifically trained on this exact inference.

\begin{tcolorbox}[colframe=red, colback=red!10, coltitle=black, top=2.5mm, bottom=2.5mm, left=1.5mm, right=1.5mm]

\scriptsize \ttfamily \textbf{QUESTION:} how much of the world's diamonds does de beers own?

\end{tcolorbox}
\vspace{-5mm}
\begin{tcolorbox}[colframe=green, colback=green!10, coltitle=black, top=1.5mm, bottom=1.5mm, left=1.5mm, right=1.5mm]

\scriptsize \ttfamily \textbf{RESPONSE:} De Beers owns 40\% of the world's diamonds.

\end{tcolorbox}
\vspace{-5mm}
\begin{tcolorbox}[colframe=orange, colback=orange!10, coltitle=black, top=2.5mm, bottom=2.5mm, left=1.5mm, right=1.5mm]

\scriptsize \ttfamily \textbf{CLAIM:} De Beers owns 40\% of the world's diamonds.

\end{tcolorbox}
\vspace{-5mm}
\begin{tcolorbox}[colframe=blue, colback=blue!10, coltitle=black, top=1.5mm, bottom=1.5mm, left=1.5mm, right=1.5mm]

\scriptsize \ttfamily \textbf{REFERENCE:} Title: Diamond Section: Industry, Gem-grade diamonds. The De Beers company, as the world's largest diamond mining company, holds a dominant position in the industry, and has done so since soon after its founding in 1888 by the British businessman Cecil Rhodes.\\ .....\\ .....\\ 
De Beers sold off the vast majority of its diamond stockpile in the late 1990s – early 2000s and the remainder largely represents working stock (diamonds that are being sorted before sale). This was well documented in the press but remains little known to the general public.

\end{tcolorbox}

\begin{figure}
\centering
\begin{tcolorbox}[colframe=brown, colback=yellow!10, coltitle=black, top=2.5mm, bottom=2.5mm, left=1.5mm, right=1.5mm]

\ttfamily Answer the question with ONLY a `YES' or `NO.' Does the REFERENCE entail the CLAIM?

\end{tcolorbox}
\caption{For our zero-shot experiments, we used this prompt template to query the LLM for determining whether the REFERENCE entails the CLAIM.}
\label{fig:prompt-template}
\end{figure}

In our problem formulation, we task the LLM with a textual entailment problem by utilizing the prompt outlined in \cref{fig:prompt-template}. This process involves evaluating the relationship between the given claim and its associated references, as defined in AttributionBench.

\subsection{Attention-based attribution}

Given the computational limitations, we designed experiments using a single LLM, specifically the \texttt{flan-t5-small} model, to analyze attention layers in addressing the attribution task. 

\textbf{Experimental Setup:} We utilized the attention weights from each layer as input to a fully connected layer for binary attribution classification. We did this for all 12 layers.


\begin{table*}[h!]
\centering
\scriptsize
\resizebox{\textwidth}{!}{%
\begin{tabular}{ll|ccc|ccc|ccc|c}\toprule 
\textbf{Setting } &\textbf{Model (Size)} &\multicolumn{3}{c}{\textbf{BEGIN (\# = 436)}} &\multicolumn{3}{c}{\textbf{AttrScore-GenSearch (\# 162)}} &\multicolumn{3}{c}{\textbf{HAGRID (\# = 1013)}} &\textbf{OOD-Avg.} \\\toprule
& &\textbf{F1 ↑} &\textbf{FP ↓ } &\textbf{FN ↓ } &\textbf{F1 ↑} &\textbf{FP ↓ } &\textbf{FN ↓ } &\textbf{F1 ↑} &\textbf{FP ↓ } &\textbf{FN ↓ } &\textbf{F1 ↑} \\\toprule

&\textbf{FLAN-T5 (770M)}  
& \rowhighlight 79.6 & \rowhighlight 9.2 & \rowhighlight 11.2 & \rowhighlight 80.8 & \rowhighlight 6.2 & \rowhighlight 13 & \rowhighlight 75.9 & \rowhighlight 13.1 & \rowhighlight 10.9 & \rowhighlight 78.8 \\ 

&\textbf{FLAN-T5 (3B)} 
& 80.2 & 13.3 & 6.4 & 82 & 6.2 & 11.7 & 79 & 16.9 & 3.8 & 80.4 \\ 

&\textbf{AttrScore-FLAN-T5 (3B)} 
& \rowhighlight 78.9 & \rowhighlight 17.7 & \rowhighlight 3 & \rowhighlight 76.3 & \rowhighlight 16.7 & \rowhighlight 6.8 & \rowhighlight 68.6 & \rowhighlight 26.9 & \rowhighlight 2.6 & \rowhighlight 74.6 \\ 

&\textbf{FLAN-T5 (11B)} 
& 72.3 & 25 & 1.1 & 78.1 & 16.7 & 4.9 & 64.5 & 30.6 & 2 & 71.6 \\ 

&\textbf{T5-XXL-TRUE (11B)} 
& \rowhighlight \textbf{86.4} & \rowhighlight 4.8 & \rowhighlight 8.7 & \rowhighlight 76.4 & \rowhighlight 2.5 & \rowhighlight 20.4 & \rowhighlight 78.6 & \rowhighlight 14.4 & \rowhighlight 6.8 & \rowhighlight 80.5 \\ 

\textbf{Zero-shot \citet{li-etal-2024-attributionbench}} &\textbf{Flan-UL2 (20B)} 
& 82.2 & 13.1 & 4.6 & 87.7 & 5.6 & 6.8 & 73.9 & 21.4 & 3.9 & 81.3 \\ 

&\textbf{AttrScore-Alpaca (7B)} 
& \rowhighlight 75.9 & \rowhighlight 20.4 & \rowhighlight 3 & \rowhighlight 82.1 & \rowhighlight 6.8 & \rowhighlight 11.1 & \rowhighlight 73.9 & \rowhighlight 19.9 & \rowhighlight 5.6 & \rowhighlight 77.3 \\  

&\textbf{GPT-3.5 (w/o CoT)} 
& 79.4 & 15.8 & 4.4 & 76.7 & 18.5 & 4.3 & 70.1 & 25.2 & 2.8 & 75.4 \\  

&\textbf{GPT-3.5 (w/ CoT)} 
& \rowhighlight 77.6 & \rowhighlight 14.9 & \rowhighlight 7.3 & \rowhighlight 82.1 & \rowhighlight 11.1 & \rowhighlight 6.8 & \rowhighlight 74 & \rowhighlight 19.7 & \rowhighlight 5.1 & \rowhighlight 77.9 \\  

&\textbf{GPT-4 (w/o CoT)} 
& 77.5 & 19.7 & 2.1 & 84.3 & 14.2 & 1.2 & 72.1 & 23.9 & 2.8 & 78 \\  

&\textbf{GPT-4 (w/ CoT)} 
& \rowhighlight 77.5 & \rowhighlight 18.3 & \rowhighlight 3.7 & \rowhighlight 83.3 & \rowhighlight 8 & \rowhighlight 8.6 & \rowhighlight 75.9 & \rowhighlight 18.5 & \rowhighlight 5.2 & \rowhighlight 78.9 \\  

\cmidrule{1-12}

\multirow{2}{*}{\textbf{Our Zero-shot}} &\textbf{gpt4-o (05-13-2024)} 
&  79.69 &  42.66 & 5.5 & \textbf{88.24} &  17.28 &  7.4 &  76.54 &  42.37 &  14.41 &  81.48 \\ 

&\textbf{flan-ul2 (20B)} 
& \rowhighlight 81.55 & \rowhighlight 32.56 & \rowhighlight 8.71 & \rowhighlight 88.05 & \rowhighlight 9.87 & \rowhighlight 13.5 & \rowhighlight \textbf{80.71} & \rowhighlight 42.79 & \rowhighlight 6.36 & \rowhighlight \textbf{83.43} \\

\bottomrule 
\end{tabular}%
}
\caption{We evaluate our zero-shot approach against the AttributionBench. Results highlighted in \textbf{bold} denote the highest performance. Our method performs better than existing approaches on the AttrScore-GenSearch and HAGRID sub-datasets. The out-of-distribution (OOD) average achieved with our approach utilizing the \texttt{flan-ul2} model is the highest, reaching a value of \textbf{83.43}.}
\label{tab:results-ood}
\end{table*}

\section{Results and Analysis}

In the initial phase of our evaluation of the attribution task, we conduct zero-shot experiments. The framework presented in AttributionBench is divided into two key components: in-distribution (ID) and out-of-distribution (OOD) sampling of the dataset. In their experimental setup, AttributionBench employs F1 score, False Positive (FP), and False Negative (FN) rates as evaluation metrics. Consistent with their methodology, we adopt the same metrics - \textbf{F1}, \textbf{FP}, and \textbf{FN} - for the evaluation in this study.

\subsection{Evaluation Metrics}
\textbf{F1:} The F1 score is a metric used to evaluate the performance of a classification model, specifically its balance between precision and recall. 

\noindent\textbf{FP:} The False Positive Rate (FP) is a measure used to evaluate the performance of a classification model, specifically in binary classification tasks. It quantifies the proportion of negative instances that are incorrectly classified as positive. 

\noindent\textbf{FN:} The False Negative (FN) is a metric used to evaluate the performance of classification models. It represents the proportion of actual positive instances incorrectly classified as negative.

\subsection{Zero-shot}

In this zero-shot setup, we formulate the attribution binary classification task as a simple \emph{textual entailment} problem. To do so, we prompt the LLM using the template shown in \cref{fig:prompt-template}. We compare our zero-shot method with the baseline zero-shot approach given in \citet{li-etal-2024-attributionbench}. With this simple question, we outperform the baselines in both ID and OOD sets.

We present our zero-shot experimental results in \cref{tab:results-id} for ID data distribution. We mainly used two LLMs: \texttt{gpt4-o} \citet{achiam2023gpt} and \texttt{flan-ul2} \citet{raffel2020exploring}. We observe that \texttt{flan-ul2} performs better with F1 accuracy metrics in Stanford-GenSearch and the LFQA sub-dataset. The best ID-average (\texttt{flan-ul2}) = \textbf{73.8}.

Similar to the results observed for in-distribution (ID) data, the highest-performing model for out-of-distribution (OOD) tasks, as presented in \cref{tab:results-ood}, is \texttt{flan-ul2}, specifically for the AttrScore-GenSearch and HAGRID sub-datasets. When evaluating the OOD performance, our approach, leveraging the \texttt{flan-ul2} model, achieves the highest average score, reaching an impressive value of \textbf{83.43}. This demonstrates the robustness and superior generalization capability of the \texttt{flan-ul2} model across both ID and OOD settings.

\subsection{Using Attention layers}

Preliminary results comparing zero-shot and varying attention layers on the LFQA attribution subset are presented in \cref{tab:atten_res}. We present layer-wise performance results for all three evaluation metrics. Although the results are mixed, the F1 scores generally outperform the baseline across nearly all layers, except for layers 4 and 8 to 11. Additionally, lower values of false positives (FP) and false negatives (FN) compared to the zero-shot baseline suggest improved performance.

\begin{table}[h!]
\centering
\scriptsize
\begin{tabular}{lccc}\toprule
\multicolumn{4}{c}{\textbf{LFQA (\#=168) }} \\\toprule
\textbf{} &\textbf{F1 ↑} &\textbf{FP ↓ } &\textbf{FN ↓ } \\ \hline
\textbf{Our Zero-shot} &20 &17.85 &86.9 \\ \hline
\multicolumn{4}{c}{\textbf{using attention}} \\ \hline
\textbf{layer 1} & \rowhighlight 66.67 &  \rowhighlight 100 &  \rowhighlight 0 \\
\textbf{layer 2} &66.93 &98.8 &0 \\
\textbf{layer 3} &  \rowhighlight  66.67 &  \rowhighlight  100 &  \rowhighlight 0 \\
\textbf{layer 4} &0 &0 &100 \\
\textbf{layer 5} &\rowhighlight  66.13 &  \rowhighlight  100 &  \rowhighlight  1.19 \\
\textbf{layer 6} &66.13 &100 &1.19 \\
\textbf{layer 7} &  \rowhighlight  65.6 &  \rowhighlight  100 &  \rowhighlight  2.38 \\
\textbf{layer 8} &10.31 &9.52 &94.04 \\
\textbf{layer 9} &  \rowhighlight  2.35 &  \rowhighlight  0 &  \rowhighlight  98.8 \\
\textbf{layer 10} &0 &0 &100 \\
\textbf{layer 11} &  \rowhighlight  66.67 &  \rowhighlight  100 &  \rowhighlight  0 \\
\textbf{layer 12} &66.93 &98.8 &0 \\
\bottomrule
\end{tabular}
\caption{With balanced classes (84 each Class 0/1) using \texttt{flan-t5-small}, F1 scores exceed the baseline across most layers, except 4 and 8–11, indicating improved performance, further supported by reduced false positives and false negatives.}
\label{tab:atten_res}
\end{table}

\section{Conclusion and Future Work}

In this paper, we conducted zero-shot experiments on AttributionBench to assess the performance of textual entailment-based approaches for attribution tasks. Our findings show that even without fine-tuning, a simple zero-shot textual entailment approach outperforms the existing baseline in both in-distribution and out-of-distribution settings. Notably, \texttt{flan-ul2} demonstrated strong performance across these scenarios, underscoring its robustness and suitability for such tasks. We also preliminary analyzed attention layer behavior using the smaller \texttt{flan-t5-small} model. The results suggest that attention mechanisms could provide valuable insights for improving attribution performance.

We plan to overcome computational limitations for future work by conducting fine-tuning experiments. We aim to use more advanced LLMs to perform a deeper analysis of attention layers. This could provide further actionable insights to refine performance and yield more robust findings.


\section{Limitations}

\paragraph{Limitation 1:} Although fine-tuning could enhance the results beyond zero-shot, it comes with additional computational overhead. Therefore, we restricted our experiments to zero-shot settings in this paper and demonstrated how a straightforward zero-shot textual entailment approach can further improve performance.

\noindent \paragraph{Limitation 2:} Regarding exploring attention mechanisms to enhance the performance of the attribution task, we were similarly restricted by computational limitations. Consequently, we could not utilize computationally demanding models for this analysis. Instead, the experiments were conducted using a lightweight model, \texttt{flan-t5-small}. 

\bibliography{custom}

\begin{thebibliography}{12}
\providecommand{\natexlab}[1]{#1}

\bibitem[{Achiam et~al.(2023)Achiam, Adler, Agarwal, Ahmad, Akkaya, Aleman, Almeida, Altenschmidt, Altman, Anadkat et~al.}]{achiam2023gpt}
Josh Achiam, Steven Adler, Sandhini Agarwal, Lama Ahmad, Ilge Akkaya, Florencia~Leoni Aleman, Diogo Almeida, Janko Altenschmidt, Sam Altman, Shyamal Anadkat, et~al. 2023.
\newblock \href {https://arxiv.org/abs/2303.08774} {Gpt-4 technical report}.
\newblock \emph{arXiv preprint arXiv:2303.08774}.

\bibitem[{Adewumi et~al.(2024)Adewumi, Habib, Alkhaled, and Barney}]{adewumi2024limitations}
Tosin Adewumi, Nudrat Habib, Lama Alkhaled, and Elisa Barney. 2024.
\newblock \href {https://arxiv.org/abs/2404.04631} {On the limitations of large language models (llms): False attribution}.
\newblock \emph{Preprint}, arXiv:2404.04631.

\bibitem[{Bohnet et~al.(2023)Bohnet, Tran, Verga, Aharoni, Andor, Soares, Ciaramita, Eisenstein, Ganchev, Herzig, Hui, Kwiatkowski, Ma, Ni, Saralegui, Schuster, Cohen, Collins, Das, Metzler, Petrov, and Webster}]{bohnet2023attributed}
Bernd Bohnet, Vinh~Q. Tran, Pat Verga, Roee Aharoni, Daniel Andor, Livio~Baldini Soares, Massimiliano Ciaramita, Jacob Eisenstein, Kuzman Ganchev, Jonathan Herzig, Kai Hui, Tom Kwiatkowski, Ji~Ma, Jianmo Ni, Lierni~Sestorain Saralegui, Tal Schuster, William~W. Cohen, Michael Collins, Dipanjan Das, Donald Metzler, Slav Petrov, and Kellie Webster. 2023.
\newblock \href {https://arxiv.org/abs/2212.08037} {Attributed question answering: Evaluation and modeling for attributed large language models}.
\newblock \emph{Preprint}, arXiv:2212.08037.

\bibitem[{Khalifa et~al.(2024)Khalifa, Wadden, Strubell, Lee, Wang, Beltagy, and Peng}]{khalifa2024sourceaware}
Muhammad Khalifa, David Wadden, Emma Strubell, Honglak Lee, Lu~Wang, Iz~Beltagy, and Hao Peng. 2024.
\newblock \href {https://openreview.net/forum?id=UPyWLwciYz} {Source-aware training enables knowledge attribution in language models}.
\newblock In \emph{First Conference on Language Modeling}.

\bibitem[{Lee et~al.(2024)Lee, Wang, Chakravarthy, Helbling, Peng, Phute, Chau, and Kahng}]{lee2024llm}
Seongmin Lee, Zijie~J. Wang, Aishwarya Chakravarthy, Alec Helbling, ShengYun Peng, Mansi Phute, Duen~Horng Chau, and Minsuk Kahng. 2024.
\newblock \href {https://arxiv.org/abs/2404.01361} {Llm attributor: Interactive visual attribution for llm generation}.
\newblock \emph{Preprint}, arXiv:2404.01361.

\bibitem[{Li et~al.(2023)Li, Sun, Hu, Liu, Chen, Hu, Wu, and Zhang}]{li2023survey}
Dongfang Li, Zetian Sun, Xinshuo Hu, Zhenyu Liu, Ziyang Chen, Baotian Hu, Aiguo Wu, and Min Zhang. 2023.
\newblock \href {https://arxiv.org/abs/2311.03731} {A survey of large language models attribution}.
\newblock \emph{Preprint}, arXiv:2311.03731.

\bibitem[{Li et~al.(2024)Li, Yue, Liao, and Sun}]{li-etal-2024-attributionbench}
Yifei Li, Xiang Yue, Zeyi Liao, and Huan Sun. 2024.
\newblock \href {https://doi.org/10.18653/v1/2024.findings-acl.886} {{A}ttribution{B}ench: How hard is automatic attribution evaluation?}
\newblock In \emph{Findings of the Association for Computational Linguistics: ACL 2024}, pages 14919--14935, Bangkok, Thailand. Association for Computational Linguistics.

\bibitem[{Miglani et~al.(2023)Miglani, Yang, Markosyan, Garcia-Olano, and Kokhlikyan}]{miglani-etal-2023-using}
Vivek Miglani, Aobo Yang, Aram Markosyan, Diego Garcia-Olano, and Narine Kokhlikyan. 2023.
\newblock \href {https://doi.org/10.18653/v1/2023.nlposs-1.19} {Using captum to explain generative language models}.
\newblock In \emph{Proceedings of the 3rd Workshop for Natural Language Processing Open Source Software (NLP-OSS 2023)}, pages 165--173, Singapore. Association for Computational Linguistics.

\bibitem[{Pasunuru et~al.(2023)Pasunuru, Sinha, Aghajanyan, YU, Wang, Bikel, Zettlemoyer, Fazel-Zarandi, and Celikyilmaz}]{pasunuru2023eliciting}
Ramakanth Pasunuru, Koustuv Sinha, Armen Aghajanyan, LILI YU, Tianlu Wang, Daniel~M Bikel, Luke Zettlemoyer, Maryam Fazel-Zarandi, and Asli Celikyilmaz. 2023.
\newblock Eliciting attributions from llms with minimal supervision.

\bibitem[{Raffel et~al.(2020)Raffel, Shazeer, Roberts, Lee, Narang, Matena, Zhou, Li, and Liu}]{raffel2020exploring}
Colin Raffel, Noam Shazeer, Adam Roberts, Katherine Lee, Sharan Narang, Michael Matena, Yanqi Zhou, Wei Li, and Peter~J Liu. 2020.
\newblock \href {https://jmlr.org/papers/volume21/20-074/20-074.pdf} {Exploring the limits of transfer learning with a unified text-to-text transformer}.
\newblock \emph{Journal of machine learning research}, 21(140):1--67.

\bibitem[{Yue et~al.(2023)Yue, Wang, Chen, Zhang, Su, and Sun}]{yue-etal-2023-automatic}
Xiang Yue, Boshi Wang, Ziru Chen, Kai Zhang, Yu~Su, and Huan Sun. 2023.
\newblock \href {https://doi.org/10.18653/v1/2023.findings-emnlp.307} {Automatic evaluation of attribution by large language models}.
\newblock In \emph{Findings of the Association for Computational Linguistics: EMNLP 2023}, pages 4615--4635, Singapore. Association for Computational Linguistics.

\bibitem[{Zhou et~al.(2024)Zhou, Adel, Schuff, and Vu}]{zhou-etal-2024-explaining}
Wei Zhou, Heike Adel, Hendrik Schuff, and Ngoc~Thang Vu. 2024.
\newblock \href {https://aclanthology.org/2024.lrec-main.600/} {Explaining pre-trained language models with attribution scores: An analysis in low-resource settings}.
\newblock In \emph{Proceedings of the 2024 Joint International Conference on Computational Linguistics, Language Resources and Evaluation (LREC-COLING 2024)}, pages 6867--6875, Torino, Italia. ELRA and ICCL.

\end{thebibliography}

\end{document}